\documentclass[twocolumn,preprintnumbers,amsmath,amssymb,aps,pre,longbibliography]{revtex4-1}
\usepackage{graphicx}
\begin{document}

\title{
Directional Locking Effects for Active Matter Particles Coupled to a Periodic Substrate 
} 
\author{
C. Reichhardt and C. J. O. Reichhardt 
} 
\affiliation{
Theoretical Division and Center for Nonlinear Studies,
Los Alamos National Laboratory, Los Alamos, New Mexico 87545, USA\\ 
} 

\date{\today}
\begin{abstract}
Directional locking occurs when a particle moving over a periodic substrate   
becomes constrained to travel along certain
substrate symmetry directions.
Such locking effects
arise for colloids and
superconducting vortices moving over ordered substrates when
the direction of the external drive is varied.
Here we study the directional locking of
run-and-tumble active matter particles interacting
with a periodic array of obstacles.
In the absence of an external biasing force,
we find that the active particle motion locks to various symmetry 
directions of the substrate
when the run time between tumbles is large.
The number of possible locking directions depends on the array density
and on the relative sizes of the particles and the obstacles.
For a square array of large obstacles,
the active particle only locks to the $x$, $y$, and
$45^{\circ}$ directions, while for smaller obstacles,
the number of locking angles increases.
Each locking angle satisfies
$\theta = \arctan(p/q)$, where $p$ and $q$ are integers, and the angle
of motion can be measured using
the ratio of the velocities or the velocity distributions
in the $x$ and $y$ directions.
When a biasing driving force is applied,
the directional locking behavior is affected by
the ratio of the self-propulsion force to the biasing force.
For large biasing, the behavior resembles 
that found for directional locking in passive systems.
For large obstacles under biased driving,
a trapping behavior occurs that is non-monotonic
as a function
of increasing run length or increasing self-propulsion force,
and the trapping diminishes when
the run length is sufficiently large.
\end{abstract}
\maketitle

\section{Introduction}

In active matter systems, particles move under a self-propulsion force.
Examples of active systems include swimming 
bacteria or self-driven colloids, and 
the activity often
takes the form of driven diffusion
or run-and-tumble dynamics \cite{Marchetti13,Bechinger16}.
In the absence of a substrate, these systems exhibit a rich variety of
collective effects such as motility induced phase separation
\cite{Fily12,Redner13,Cates15,Palacci13,Buttinoni13}.
When barriers or a substrate is added,
a number of behaviors arise
that are distinct from
what is observed for purely Brownian particles
\cite{Yang14,Mallory14,Takatori14,Ray14,Ni15,Solon15,Speck20}.
For example, in
active matter ratchet effects,
active particles interacting with some form of
asymmetric barrier or obstacles undergo directed flow in the absence of
dc driving,
while for equilibrium or Brownian particles  
such ratchet effects are absent
\cite{Wan08,Nikola16,Reichhardt17a,Berdakin13,Lozano16}.

A growing number of studies
have examined active matter systems coupled to a complex environment such as   
randomly disordered sites
\cite{Chepizhko15,Morin17,Sandor17a,Morin17a,Zeitz17,Reichhardt18c,Bertrand18,Chepizhko19,Bhattacharjee19,Zeitz17,Chardac20}  or
a periodic array of obstacles
\cite{Volpe11,Quint15,AlonsoMatilla19,Zeitz17,Jakuszeit19,Pattanayak19,Schakenraad20,Ribeiro20,Yazdi20,BrunCosmeBruny20}.
Extensive studies of passive particles
on a periodic array of obstacles 
under diffusion
\cite{Brunner02,Mangold03,Reichhardt05,Lacasta04,HerreraVelarde09a}
or an external drive \cite{Reichhardt17,Bohlein12,Vanossi12,McDermott13a}
show that depinning phenomena, sliding phases, and
different types of dynamical pattern formation
appear when collective effects become important.
Passive particles driven over
a periodic substrate can undergo
directional locking in which
the motion
becomes locked to certain symmetries of the lattice.
Here, the direction of particle motion
does not change smoothly 
as the angle of the external drive is rotated with respect to the
substrate,
but remains fixed for finite intervals of drive angle,
producing
steps in a plot of the direction of 
particle motion versus drive direction 
\cite{Reichhardt99,Wiersig01,Korda02,Silhanek03,Gopinathan04,Balvin09,Koplik10,Reichhardt12,Li19,Tierno19,Cao19,Trillitzsch18,Stoop20}. 
For a particle moving on a square lattice,  
the directional locking occurs
when the particle moves $p$ lattice constants
in one direction and
$q$ lattice constants in the perpendicular direction,
giving locking steps centered
at angles of $\theta = \arctan(p/q)$. 
For example, $p/q = 0$ and $p/q = 1/1$ correspond to locking
at $0^\circ$ or $+45^\circ$, while
many other locking phases
can appear at $|p/q|=1/4$, 1/3, 1/2, 2/3, 3/4, 2/1, 3/1, and so forth. 
The number of
locking steps depends on the
relative radius of the obstacles and the particle.
Small obstacles with a fixed lattice constant produce
a devil's staircase hierarchy of locking
steps, with the smallest values of $|p/q|$ giving the largest step widths. 
If the substrate is a triangular lattice, directional 
locking occurs at a different set of angles,
including $30^\circ$ and $60^\circ$ \cite{Stoop20,Reichhardt04b}.

Directional locking
effects have been studied for vortices in type-II superconductors
moving over periodic pinning arrays
\cite{Reichhardt99,Silhanek03,Reichhardt12}, classical
electrons moving over antidot lattices \cite{Wiersig01}, 
atomic motion on surfaces \cite{Trillitzsch18},
and magnetic skyrmions driven over periodic landscapes
\cite{Reichhardt15a,Feilhauer19,Vizarim20}. 
The most studied directional locking system
is colloids moving over either optical trap arrays
or periodic arrays of posts.
Here, the colloidal motion becomes locked to different symmetry
directions of the substrate as either
the effective biasing drive changes direction or
the substrate itself is rotated under a fixed driving direction
\cite{Korda02,Gopinathan04,Balvin09,Reichhardt12,Tierno19,Stoop20,Reichhardt04b}. 
Similar locking effects can even arise
for particles moving over quaisperiodic substrates
\cite{Reichhardt11,Bohlein12a}. 
The strength of the locking or width of the
locking steps depends strongly on the properties of the particles
such as their size, shape, and particle-substrate 
interactions, and as a result,
directional locking acts as a powerful method
for sorting mixtures of different particle species.
When one species locks to a symmetry direction and
another species either does not lock or locks to a different
symmetry direction, the species separate laterally over time.
Such fractionation effects have been demonstrated for a number
of colloidal \cite{MacDonald03,Pelton04,Ladavac04,Lacasta05,Roichman07a} 
and other soft matter systems
\cite{Huang04,Long08,Speer10,Risbud14,Wunsch16,Tran17,Li18}.     

An open question is whether directional locking 
also occurs for active matter systems coupled to a periodic substrate.
Such systems could potentially exhibit
directional locking even in the absence of an externally applied
biasing field. 
Volpe {\it et al.} experimentally examined active matter particles
moving over a triangular lattice of posts under a biasing drive
for varied activity, and found that the
particles can lock to the symmetry directions of the
substrate even when the external driving is not aligned with 
these symmetry directions \cite{Volpe11}.
More recently,
Brun-Cosme-Bruny {\it et al.} examined swimming micro-algae driven
phototactically through a square array of obstacles, 
and found that the motion locks to certain symmetry
angles \cite{BrunCosmeBruny20}. These results suggest that a
variety of directional locking effects
should also occur for active matter systems with and without biasing drives.   

In this work we examine run-and-tumble active matter disks
moving through a square array of obstacles. In the absence of a biasing field,
we find that for small run lengths the system behavior is
close to the diffusive limit and the particles explore the
background in a uniform fashion; however, for long
run lengths the particle motion becomes
locked to specific symmetry directions of the obstacle lattice.
The number of possible locking directions
depends strongly
on the size of the obstacles.
For large obstacles, the particles lock along
$\theta=0^\circ$, $45^\circ$, and $90^\circ$,
while for smaller obstacles,
additional locking directions appear with
$\theta =  \arctan(p/q)$ for integer $p$ and $q$. 
The locking can be measured using
the particle velocity distributions,
which show peaks on the locking steps.
As the obstacle radius
increases, we find
an increasing 
probability for the particles
to become trapped by the obstacles. 
When there is an additional applied driving force,
the net drift velocity in the driving direction
has a non-monotonic 
behavior as a function of
the run length and the ratio of the motor force to the driving force.
When the drive direction is rotated
from $0^\circ$ to $90^\circ$,
we observe a series of locking steps
which are more pronounced
for lower run lengths and smaller motor forces.
If the motor force is made very large, the locking steps
disappear, but they can be restored by
increasing the magnitude of the driving force.
For large obstacles which produce a clogged state
in the passive particle limit,
we find that the activity can induce motion along
certain substrate symmetry directions,
producing a
nonmonotonic mobility that depends on the direction of drive
relative to the symmetry directions of the
substrate.  

\section{Simulation}

We consider $N_d$ active run-and-tumble disks
interacting with a square array of obstacles 
in a two-dimensional system of size $L \times L$.
The radius of each obstacle, modeled as a short-range repulsive disk,
is $r_{\rm obs}$, and $a$ is the
obstacle lattice constant.
The active particles are also modeled as short-range repulsive disks with
radius $r_{a}$.  
The dynamics of active disk $i$ is obtained by integrating the
following overdamped equation of motion: 
\begin{equation} 
\alpha_d {\bf v}_{i}  =
{\bf F}^{dd}_{i} + {\bf F}^{m}_{i} + {\bf F}^{\rm obs}_{i} + {\bf F}^{D} .
\end{equation}
The velocity of the active 
particle is ${\bf v}_{i} = {d {\bf r}_{i}}/{dt}$, where  ${\bf r}_{i}$
is the disk position.
We set the damping constant $\alpha_d$ to $1.0$. 
The disks can interact with each other
via the disk-disk interaction force ${\bf F}^{dd}_{i}$
and with the obstacles via
the disk-obstacle force ${\bf F}^{\rm obs}$.
In general, we consider the limit $N_d=1$ of a single active particle,
so the active disk-disk interactions are not important.
The self-propulsion of the disk is produced
by the motor force ${\bf F}^{m}_i$, a constant 
force of magnitude $F_{M}$ that is applied in the randomly chosen direction
${\bf \hat{m}}_i$ for a run time of $\tau_{l}$.
After the run time has elapsed, a new motor force direction
${\bf \hat{m}^\prime}_i$ is chosen randomly, corresponding to instantaneous
reorientation, and the particle travels under the same motor force $F_M$ in
the new direction for a time $\tau_l$ before the motor force reorients
again.
We can add an external biasing force ${\bf F}^D$, which
is first taken to have a fixed orientation
along the $x$-direction with a magnitude $F_{D}$. 
We also consider the case where the magnitude $F_D$ of the drive is fixed but 
the direction gradually rotates from $\theta_D=0$ along the $x$ direction to
$\theta_D=90^\circ$ along the $y$ direction.
Here, ${\bf F}_{D} = F_{D}\cos(\theta_D){\bf \hat{x}} + F_{D}\sin(\theta_D){\bf \hat{y}}$. 
We measure the average velocity in the $x$-direction, 
$\langle V_{x}\rangle = \sum^{N_{d}}_{i= 1}{\bf v}_i\cdot {\bf \hat{x}}$,
and in the $y$-direction,
$\langle V_{y}\rangle = \sum^{N_{d}}_{i= 1}{\bf v}_i\cdot {\bf \hat{y}}$.
We also measure the net velocity
$\langle V\rangle = \sqrt{\langle V_x\rangle^{2} + \langle V_y\rangle^{2}}$.
We characterize the activity based on the distance $l_r$
a free active particle would move in the absence of obstacles during
a single run time, which we term the run length $l_{r} = F_{M}\tau_{r}$, as
well as by the ratio of the motor to the drive force,
$F_{M}/F_{D}$.
In this work we focus on the low disk density regime
where the collective effects are weak,
placing the dynamics in the single particle limit.
We fix the obstacle lattice constant $a = 4.0$ and
the active disk radius $r_{a} = 0.5$,
but vary the obstacle size, the motor force, the
run length, and the external biasing force. 

\section{Directional Locking Due to Activity}

\begin{figure}
\includegraphics[width=\columnwidth]{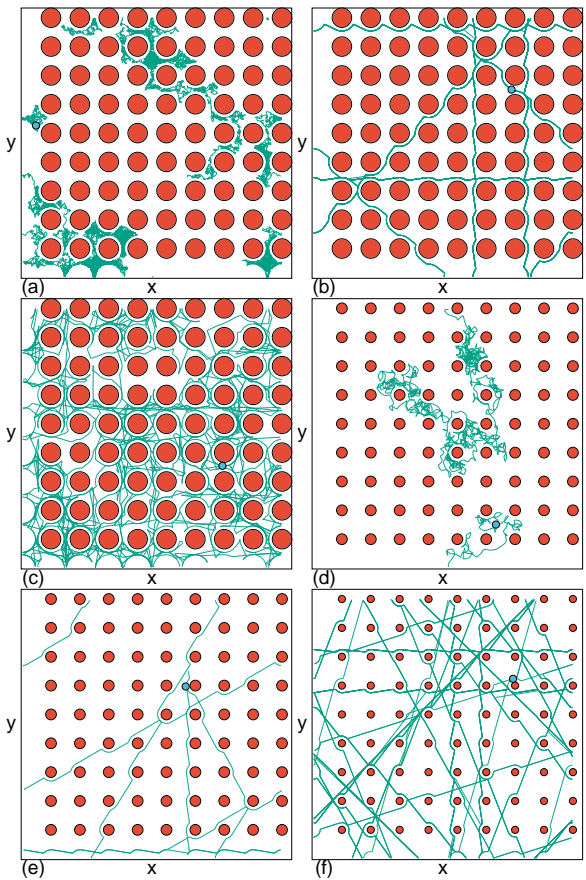}
\caption{
Illustration of the obstacle locations (red circles), the active particle
(blue circle), and the active particle trajectory (lines) in a system with
motor force $F_{M} = 0.4$ and no external drive.
(a) At obstacle radius $r_{\rm obs} = 1.35$ and run length $l_{r} = 0.3$, 
there is no directional locking.
(b) For $r_{\rm obs}=1.35$ and
$r_{l} = 80$,
the particle motion
locks to
$0^\circ$, $45^\circ$ and $90^\circ$.
(c) For
$l_{r} = 5.6$ and
$r_l=80$,
the motion is ballistic at short times and diffusive at
long times.
(d) At $r_{\rm obs} = 0.75$ and $l_{r} = 0.8$, the motion is diffusive.
(e) At $r_{\rm obs}=0.75$ and
$l_{r} = 80$, directional locking occurs with a larger number of possible
locking directions compared to panel (b).
(f)
At $r_{\rm obs} = 0.5$  and $l_r=80$, an even larger number of locking
directions appear.
}
\label{fig:1}
\end{figure}

In Fig.~\ref{fig:1} we illustrate the obstacles,
active particle, and trajectory for a 
system with $F_M=0.4$ and
no external biasing force. 
When the obstacle radius $r_{\rm obs}=1.35$ and
the run length $l_r=0.3$, as in Fig.~\ref{fig:1}(a),
the behavior is close to the Brownian limit
and the particle gradually explores the space between the obstacles while
exhibiting no directional locking.
In Fig.~\ref{fig:1}(b), where we have increased the run length to
$l_r=80$,
the particle 
moves in one-dimensional (1D) trajectories that are
locked
along the $\pm x$, $\pm y$, and $\pm 45^\circ$ directions.
At an intermediate run length of $l_r=5.6$ in
Fig.~\ref{fig:1}(c),
the particle motion is ballistic at short times but diffusive at longer times,
which allows the particle
to explore the entire system at a much faster rate compared to
the $l_r=0.1$ case.
Here, since the persistent motion spans a distance of
only about one substrate lattice constant $a$, there is no
directional locking.  
Figure~\ref{fig:1}(d) illustrates a sample with
a smaller $r_{\rm obs} = 0.75$ at $l_{r} = 0.8$, where
the motion is diffusive.
When the run length is increased to $l_r=80$, as in 
Fig.~\ref{fig:1}(e),
directional locking occurs and the particle follows 1D paths aligned not
only with $\pm x$, $\pm y$, and $\pm 45^\circ$, but also
with $\theta = \pm 26.56^\circ$ and $\theta = \pm 71^\circ$,
corresponding to a translation by $2a$ in the $x$ direction for every $a$ in
the $y$ direction or a translation by $a$ in the $x$ direction for every
$2a$ in the $y$ direction, respectively.
When the run length is large, such as for $l_r=80$, 
we find that as $r_{\rm obs}$ decreases,
the number of possible locking angles increases.
In Fig.~\ref{fig:1}(f) we show a sample with $l_r=80$ and $r_{\rm obs}=0.5$
where
the locking directions
include angles for which $|p/q| = 0$, 1/1, 1/2, 1/3, 1/4, 2, 3, and $4$.   
In general, appreciable directional locking
occurs whenever the run length is larger than $2a$. 

\begin{figure}
\includegraphics[width=\columnwidth]{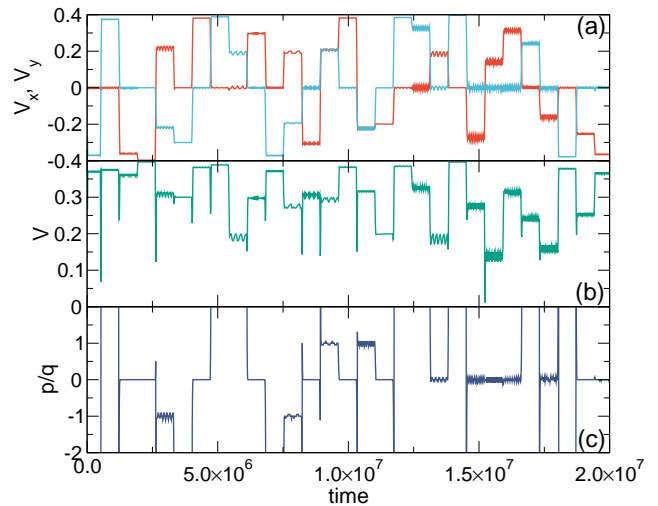}
\caption{
(a) The instantaneous velocities in the $x$-direction $V_{x}$ (orange)
and in the $y$-direction $V_{y}$ (light blue) versus time
in simulation time steps for the system
in Fig.~\ref{fig:1}(b) with $F_M=0.4$, $r_{\rm obs}=1.35$, and $l_r=80$.
(b) The corresponding total velocity
$V = \sqrt{V^2_{x} + V^{2}_{y}}$ vs time.
(c) The corresponding $p/q=V_y/V_x$ vs time.
$p/q = \pm 1.0$ for motion along $\pm 45^\circ$,
$p/q=0$ for motion along $\pm x$, and $p/q$ is $\pm$ infinite for motion
along $\pm y$.  
}
\label{fig:2}
\end{figure}

To better characterize the motion, we measure the instantaneous
velocities $V_x$ and $V_y$  in the $x$ and $y$  directions along with the ratio
$p/q=V_y/V_x$ which indicates the angle of the instantaneous motion.
In Fig.~\ref{fig:2}(a) we plot $V_{x}$ and $V_{y}$ versus time 
for the system in
Fig.~\ref{fig:1}(b) with $r_{\rm obs}=1.35$ and $l_r=80$,
where the particle moves
in either the $\pm x$, $\pm y$, or $\pm 45^\circ$ directions.
The velocity signatures are composed of steps, with
$V_x=0$ or $V_y=0$ when the velocity is locked along the $\pm y$ or $\pm x$
direction, respectively.
Since
the dynamics is overdamped,
we always have $|V_x| \leq F_M$ and $|V_y| \leq F_M$.
In Fig.~\ref{fig:2}(b) we show the
corresponding
net velocity $V  = \sqrt{V^2_{x} + V^{2}_{y}}$
versus time.
The fixed motor force might lead one to expect that $V=F_M$ at all times;
however, due to collisions with the obstacles we often find
$V<F_M$.
When the motor force direction is aligned with a symmetry direction
of the lattice, $V \approx F_M$ since few particle-obstacle collisions
occur;
however, if the motor force fails to align with any of the symmetry
directions of the lattice, periodic collisions occur and reduce the value
of $V$.
For example, if the motor force is
aligned along $15^\circ$, then
in the absence of a substrate the particle
would move at $15^\circ$ 
with $ V = F_{M}$.
When the substrate is present,
for the obstacle radius shown in Fig.~\ref{fig:2} the particle motion is 
locked to $0^\circ$ and the particle travels in the positive $x$-direction,
colliding regularly with the obstacles and moving at a reduced value
of $V$.
On the other hand, when the motor force is
aligned along $0^\circ$, which exactly matches a symmetry direction of
the substrate, no particle-obstacle collisions
occur and $V=F_M$.
As a result, locked motion along a symmetry direction such as $0^\circ$
can be associated with
a range of possible net particle velocities.
Figure~\ref{fig:2}(b) indicates that $V$ generally exhibits a
transient dip each
time the particle changes direction due to the
increased collisions that occur until the orbit stabilizes
in the new locking direction.
In Fig.~\ref{fig:2}(c) we plot the ratio
$p/q = V_{y}/V_{x}$
versus time, which takes five distinct values:
$p/q=+1.0$ for motion along $+45^\circ$,
$p/q=-1.0$ for motion along $-45^\circ$,
$p/q=0$ for motion
along $\pm x$,
positive infinite $p/q$ for motion along $+y$,
and negative infinite $p/q$ for motion along $-y$.

\begin{figure}
\includegraphics[width=\columnwidth]{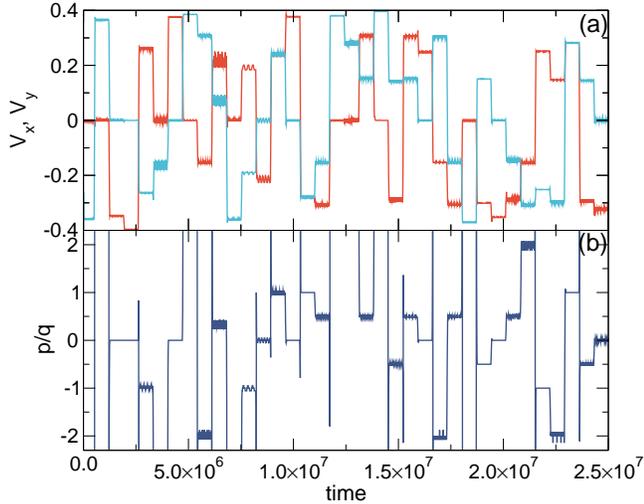}
\caption{ 
(a) Instantaneous velocities $V_{x}$ (orange) and $V_{y}$ (light blue)
vs time in simulation time steps for the system in Fig.~\ref{fig:1}(e)
with $F_M=0.4$, $r_{\rm obs} = 0.75$, and $l_r=80$. 
(b) The corresponding $p/q = V_{x}/V_{y}$ with
steps at $p/q = 0$, $\pm 1/2$, $\pm 1$, and $\pm 2$.
}
\label{fig:4}
\end{figure} 

\begin{figure}
\includegraphics[width=\columnwidth]{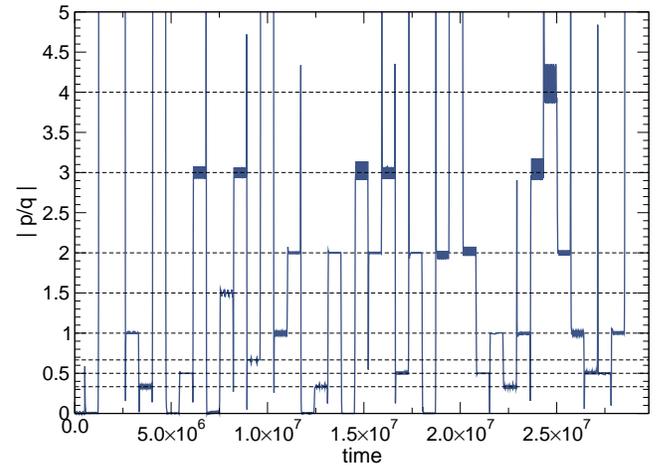}
\caption{ 
The ratio $|p/q| = |V_{y}/V_{x}|$ for a system with
$F_M=0.4$, $r_{\rm obs}=0.25$, and $l_r=80$.
Steps appear at $|p/q| = 0$, 1/3, 1/2, 2/3, 1, 3/2, 2, 3, and 4.
For longer
time intervals (not shown), additional steps occur at
$|p/q| = 1/5$, 1/4, 5/2, and 5. 
}
\label{fig:5}
\end{figure} 

In Fig.~\ref{fig:4}(a) we plot $V_{x}$ and $V_{y}$ versus
time for the system in Fig.~\ref{fig:1}(e) with $r_{\rm obs} = 0.75$
and $l_r=80$, while Fig.~\ref{fig:4}(b) shows the corresponding
$p/q=V_y/V_x$ versus time.
The velocities again undergo a series of jumps.
The smaller value of 
$r_{\rm obs}$ 
permits the active particle to access a larger number of symmetry
directions
corresponding to  $p/q = 0$, $\pm 1/2$, $\pm 1$, and $\pm 2$.
In general, when $l_{r}$ is large,
decreasing $r_{\rm obs}$ increases the number
of possible $p/q$ states, 
as shown in Fig.~\ref{fig:5}
where we plot $|p/q|$ versus time for a system with $l_r=80$ and
a smaller $r_{\rm obs}=0.25$.
Dashed lines highlight the locking steps that appear at 
$|p/q| = 0$, 1/3, 1/2, 2/3, 1, 3/2, 2, 3, and 4. For longer
times beyond what is
shown in Fig.~\ref{fig:5},
additional locking steps occur
at $|p/q|=1/5$, 1/4, 5/2, and 5. 
In general, the system spends a larger fraction of time
locked along directions that 
correspond to lower values of $p$ and $q$. 

\begin{figure}
\includegraphics[width=\columnwidth]{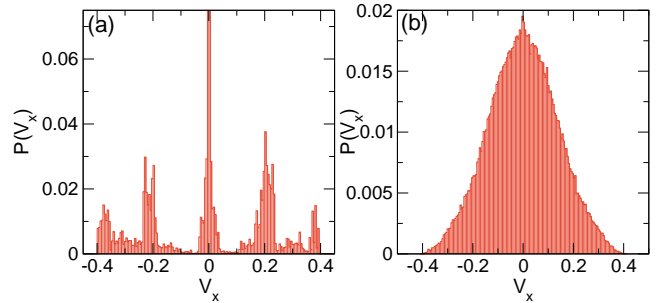}
\caption{ 
The distribution $P(V_x)$ of the instantaneous $x$ velocity 
for samples with $F_M=0.4$ and $r_{\rm obs}=1.35$.
(a) The system in Fig.~\ref{fig:1}(a) with 
$l_{r} = 80$,
where the motion is locked along $\pm x$, $\pm y$,
and $\pm 45^\circ$. 
(b) A system with 
$l_{r} = 0.2$, where the behavior is close to Brownian motion.}
\label{fig:6}
\end{figure}

For a fixed motor force and obstacle density,
we can also use the distribution $P(V_x)$ of the instantaneous
$x$ velocity 
to characterize the system. 
In Fig.~\ref{fig:6}(a) we plot
$P(V_{x})$ for the sample from Fig.~\ref{fig:1}(a) with a long run 
length of $l_{r} = 80$, where the
motion is locked to
the $\pm x$, $\pm y$, or $\pm 45^\circ$ direction, while in
Fig.~\ref{fig:6}(b) we
show a sample with $l_{r}= 0.2$ which
is close to the Brownian
limit.
When $l_r$ is small, $P(V_x)$ is nearly Gaussian, consistent with the 
expectations for a random walk. 
At large $l_r$, we find
peaks in $P(V_x)$ near $V_{x}=\pm 0.4$ which corresponds to particles
that are traveling along the $\pm x$ direction at a speed that is close
to the magnitude of the motor force.
A larger peak appears at $V_{x} = 0.0$ corresponding
to motion that is locked in the $\pm y$-direction with a finite value
of $V_y$.
The $V_x=0$ peak is not simply twice as large as either of the $V_x=\pm 0.4$
peaks, but is nearly six times higher.
This is due to a trapping effect in which the particle collides with an
obstacle and has an instantaneous velocity that is nearly zero.
For short run lengths, a particle that has become temporarily trapped due
to a collision with an obstacle quickly reorients its direction of motion
and moves away, but if the run length is long,
the particles can effectively be trapped for
an extended time.
Such active trapping by obstacles was studied
previously for run-and-tumble particles moving through random
arrays under a drift force, where the active drift velocity is
high at smaller $l_{r}$ but drops with increasing $l_{r}$
\cite{Reichhardt14,Reichhardt18}.
This trapping effect is also similar
to the active particle accumulation
that occurs along walls and corners
when the particles persistently push up against a barrier 
\cite{Yang14,Solon15,Speck20,Reichhardt17a}.
Figure~\ref{fig:6}(a) also shows
peaks in $P(V_x)$ near $V_{x} = \pm 0.22$, which
corresponds to motion along $\pm 45^\circ$.
The velocity is slightly lower than the expected value of
$|V_{x}|=F_{M}\cos(45^\circ) = 0.28$
since the motor force is generally not aligned precisely along $\pm 45^\circ$
and therefore particle-obstacle collisions occur that slow down
the particle.
Some additional subpeaks appear in $P(V_x)$, with the prominent peaks
corresponding to additional modes
of motion in which the particle collides with
a quantized number of obstacles during a given time interval.   
The distributions shown
in Fig.~\ref{fig:6} are for a single active particle.
If an increasing density of active particles is introduced,
the features in $P(V_x)$
begin to smear out
due to collisions between active particles
which lower the effective running length.

\begin{figure}
\includegraphics[width=\columnwidth]{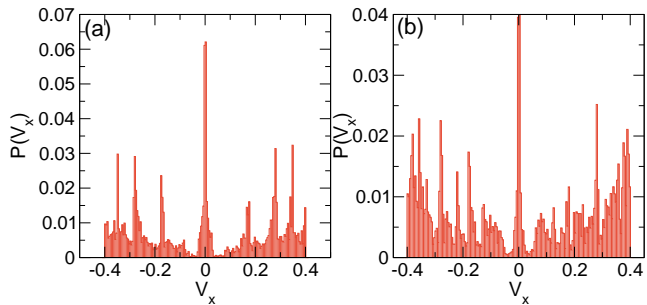}
\caption{ 
$P(V_{x})$ for samples with $F_M=0.4$ and $l_r=80$.
(a)
A sample with $r_{\rm obs} = 0.5$
shows multiple peaks at $|V_{x}| = 0$, 0.175, 0.28, and $0.35$,
corresponding to lockings of $|p/q| = 0$, 1/2, 1, and $2$. 
(b)
A sample with $r_{\rm obs} = 0.05$ has velocity peaks
corresponding to lockings with 
$p/q = 0$, 1/4, 1/3, 1/2, 2/3, 3/4, 1/1, 2/1, 3/1, 4/1, and  $5/1$.
}
\label{fig:7}
\end{figure}

In Fig.~\ref{fig:7}(a) we plot $P(V_{x})$
for
a sample with $l_r=80$ and a smaller
obstacle size of $r_{\rm obs} = 0.5$, which increases the
number of possible locking directions.
Multiple peaks appear in $P(V_x)$ centered at
$|V_{x}| = 0$, 0.175, 0.28, and $0.35$.
From the relation
$|V_{x}| = F_{M}\cos(\theta)$,
these peaks correspond to angles of motion
of $\theta=0^\circ$, $18.43^\circ$, $45^\circ$, and $63.43^\circ$,
or $|p/q| = 0$, 1/2, 1, and $2$.
In Fig.~\ref{fig:7}(b), a sample with $l_r=80$ and even smaller obstacles
with
$r_{\rm obs} = 0.05$
exhibits significantly more peaks corresponding to
$|p/q| = 0$, 1/4, 1/3, 1/2, 2/3, 3/4, 1/1, 2/1, 3/1, 4/1, and  $5/1$.        

\section{Effects of an Applied Biasing Force}

\begin{figure}
\includegraphics[width=\columnwidth]{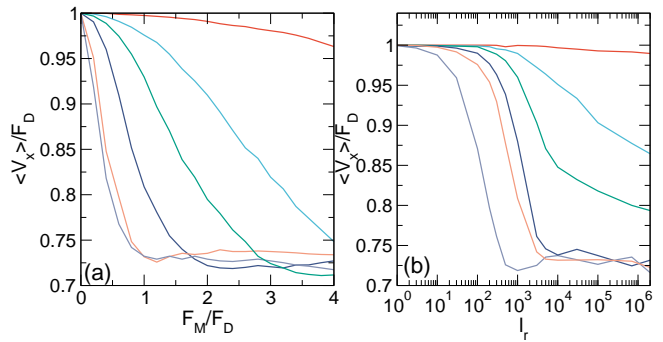}
\caption{ 
$\langle V_x\rangle/F_D$ curves for
a system with $r_{\rm obs}=1.35$ under a
fixed external biasing force $F_D=0.5$ aligned with the
positive $x$ axis.
(a) $\langle V_{x}\rangle/F_{D}$ vs $F_{M}/F_{D}$
for running lengths
$l_{r} = 0.01$, 0.1, 0.3, 1.0, 10, and 100, from top to bottom.  
(b) $\langle V_{x}\rangle/F_{D}$ vs $l_r$ for
$F_{M}/F_{D} = 0.04$, 0.2, 0.7, 1.0, and $2.4$, from top to bottom.
}
\label{fig:8}
\end{figure}

We next introduce an applied external drift force of magnitude $F_D$,
which we initially take to have a fixed magnitude of $F_D=0.5$ and
a fixed orientation along the positive $x$ direction, $\theta_D=0$.
In Fig.~\ref{fig:8}(a) we plot
$\langle V_{x}\rangle/F_{D}$ versus $F_{M}/F_{D}$
in a system 
with $r_{\rm obs} = 1.35$
at running lengths of
$l_{r} = 0.01$, 0.1, 0.3, 1.0, 10, and $100$.
When $F_M/F_D=0$, 
all of the curves coincide
at $\langle V_{x}\rangle/F_{D} = 1.0$,
indicating that the particle
can travel freely in the $x$ direction without hitting any obstacles.
As $l_{r}$ increases, $\langle V_x\rangle/F_D$ begins to decrease, with
a rapid drop in $\langle V_{x}\rangle/F_D$ occurring once
$F_{M}/F_{D} > 1.0$.
This rapid drop results when
the particle undergoes additional collisions with the obstacles and
becomes trapped behind them for periods of time.
For $l_{r} \geq 100$,
$\langle V_{x}\rangle/F_{D}$ saturates to a value close to $0.725$
when $F_{M}/F_{D} > 1.0$.    
If the obstacles are randomly placed instead of being arranged in
a lattice, a similar decrease in velocity with increasing run length
occurs,
as previously studied in a system with a biasing force
\cite{Reichhardt14,Reichhardt18}. 
Figure~\ref{fig:8}(b) shows
$\langle V_{x}\rangle/F_{D}$ versus run length for the system in
Fig.~\ref{fig:8}(a) at 
$F_{M}/F_{D} = 0.04$, 0.2, 0.7, 1.0, and $2.4$.
Here, $\langle V_{x}\rangle/F_{D} = 1.0$
at short run lengths and drops
as $l_r$ increases.
The location of the drop shifts to lower run lengths as $F_M/F_D$ increases.

\begin{figure}
\includegraphics[width=\columnwidth]{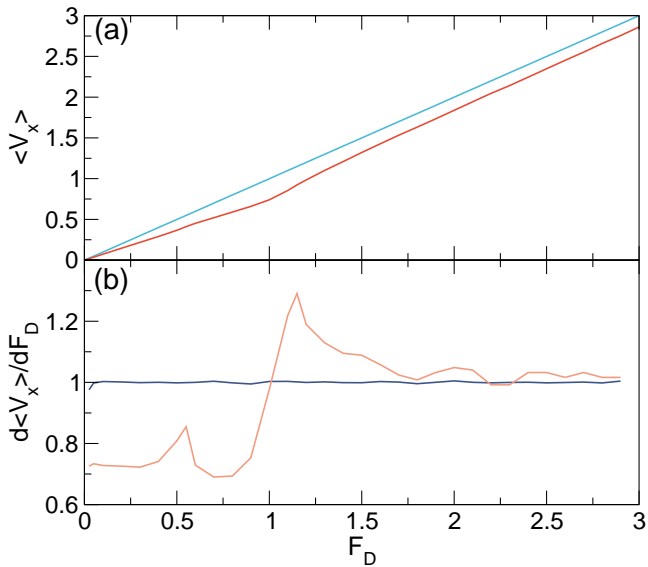}
\caption{ 
(a) $\langle V_{x}\rangle$ vs $F_{D}$
in a system with $r_{\rm obs} = 1.35$ and $x$ direction driving
at $F_{M} = 0.0$ (blue curve), showing
a linear increase in $\langle V_{x}\rangle$ with $F_{D}$,
and at 
$F_M=0.5$
and $l_{r} = 100$ (dark orange curve).  
(b) The corresponding $d\langle V_{x}\rangle/dF_{D}$ curves
showing no peak for the $F_{M} = 0.0$ system
(dark blue curve) and two peaks for the
$F_{M} = 0.4$ and $l_r=100$ system (orange curve).  
}
\label{fig:9}
\end{figure}

In Fig.~\ref{fig:9}(a) we plot $\langle V_{x}\rangle$ versus $F_{D}$
for a system with
$r_{\rm obs} =1.35$.
At $F_{M} = 0$,
the velocity increases linearly with
increasing $F_{D}$ since no collisions occur between the particles and
the obstacles.
When
$F_{M} = 0.5$
and $l_r=100$,
there is a reduction in
$\langle V_{x}\rangle$
for all values of $F_{D}$
since the inclusion of a motor force causes
particle-obstacle collisions as well as
self-trapping of the particles behind obstacles. 
Figure~\ref{fig:9}(b) shows the corresponding
$d\langle V_{x}\rangle/dF_{D}$ versus $F_{D}$ curves.
When $F_{M} = 0.0$, $d\langle V_x\rangle/dF_D$ is flat, indicating
a linear velocity-force curve,
while for the
finite motor force, two peaks appear in
$d\langle V_x\rangle/dF_D$
near $F_{D} = 0.5$ and
$F_D=1.0$.
At the first peak, $F_D \approx F_M$ and
the motion shifts from being dominated by the motor force to being
dominated by the driving force.
The large second peak appears
when $F_{D}$ becomes big enough that the particle motion is restricted
to flow entirely along the $x$ direction without any motor force-induced
jumps between adjacent flowing lanes.
Thus, the second peak indicates the occurrence of a
two-dimensional (2D) to 1D transition in
the motion.
The multiple peaks in the $d\langle V_x\rangle/dF_{D}$ curve
are similar to the behavior typically found in a system with a depinning
transition or with
a transition from effectively 2D to effectively 1D dynamics
\cite{Reichhardt17}.
For large $F_D$ values above the second peak in $d\langle V_x\rangle/dF_D$,
the system behaves in a non-active manner.       

\begin{figure}
\includegraphics[width=\columnwidth]{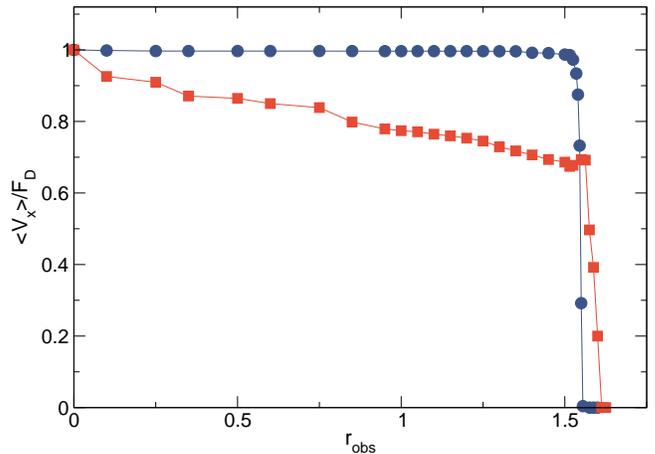}
\caption{ 
$\langle V_{x}\rangle/F_{D}$ vs $r_{\rm obs}$ for a system
with $F_M=0.5$ and $x$ direction driving of $F_D=0.2$
at $l_{r} = 0.01$ (blue circles) and $l_{r} = 10$ (orange squares).
The larger run length activity
lowers the overall mobility, but also increases
the value of $r_{\rm obs}$ at which
$\langle V_{x}\rangle/F_{D}$ drops
to zero.
}
\label{fig:10}
\end{figure}

In Fig.~\ref{fig:10} we plot $\langle V_{x}\rangle/F_{D}$ versus
$r_{\rm obs}$ for a system with
$F_{M} = 0.5$ and an $x$ direction driving force of $F_{D} = 0.2$
at both a
small run length of $l_{r} = 0.01$ and
a larger run length of $l_{r} = 10$.
For the small run length, the particle velocity matches the driving velocity
over most of the range of $r_{\rm obs}$. 
Only near $r_{\rm obs} = 1.53$ does the velocity begin to decrease precipitously
until the particle becomes
trapped with $\langle V_{x}\rangle = 0$ for $r_{\rm obs} > 1.57$.
In contrast, for the long run length $\langle V_{x}\rangle/F_{D}$
decreases with increasing $r_{\rm obs}$
due to the self trapping effects;
however,
the velocity
does not drop completely to zero until $r_{\rm obs} > 1.6$. 
In general, increased activity in the form of an increased run length
reduces the mobility through the system; however,
if strong disorder is introduced, there can also be regimes where the
higher activity increases the mobility.
If the particle radius
$r_{a}$ is increased, the curves for both run lengths shift to
lower velocities,
while the value of $r_{\rm obs}$ at which the velocity drops to zero
decreases as $r_{\rm obs} - r_{a}$. 
  
\section{Directional Locking For Varied Drive Directions}

\begin{figure}
\includegraphics[width=\columnwidth]{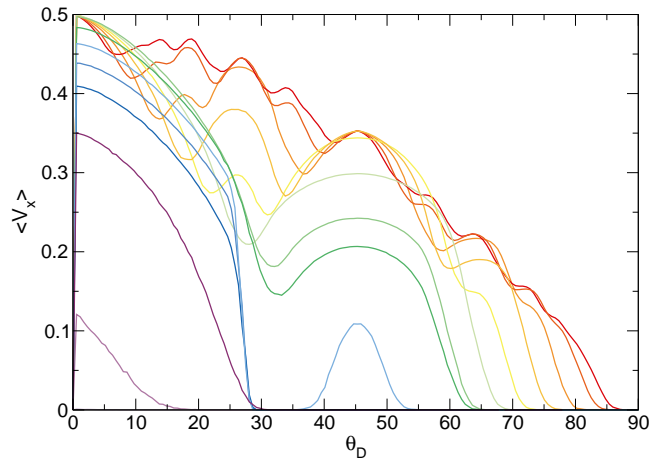}
\caption{ 
$\langle V_{x}\rangle$ vs $\theta_D$ for a system
with $F_M=0.5$ and $l_r=0.01$
under a drive
of $F_D=0.5$ applied along the $\theta_D$ direction  
at $r_{\rm obs} =  0.1$, 0.25, 0.5, 0.75, 1.0, 1.25, 1.35, 1.5,  1.55,
1.5625, 1.57, 1.575, 1.58, 1.5875, and $1.5925$,
from top (red) to bottom (purple).   
}
\label{fig:11}
\end{figure}

We next
apply a finite driving force of fixed magnitude
and rotate its direction
from $\theta_D=0^\circ$ to $\theta_D = 90^\circ$.
In Fig.~\ref{fig:11} we
plot $\langle V_{x}\rangle$ versus $\theta_D$
for a system with $F_D=0.5$, $F_{M} = 0.5$, and $l_{r} = 0.01$ at
$r_{\rm obs} =  0.1$, 0.25, 0.5, 0.75, 1.0, 1.25, 1.35, 1.5,  1.55, 1.5625,
1.57, 1.575, 1.58, 1.5875, and $1.5925$.
The velocity does not vary monotonically with $\theta_D$
but shows a series of rounded peaks
which correspond to directional locking
that would be accompanied by a series of steps in the
$p/q=\langle V_{y}\rangle/\langle V_{x}\rangle$ curves.
Note that directional locking in the non-active limit
$l_{r}  = 0.0$
was studied in detail elsewhere \cite{Reichhardt20}.
For small $r_{\rm obs} = 0.1$, we find steps with
$p/q = 0$, 1/5, 1/4, 1/3, 1/2, 2/3, 3/4, 1/1, 4/3, 3/2, 2, 3, 4, and $5$.
As $r_{\rm obs}$ increases at small $l_r$, the widths of the higher 
order steps diminish
while the steps with $p/q=0$, 1/2, 1/1, 2/1,
and $y$-direction locking increase in size as shown in Fig.~\ref{fig:11}.
At $r_{\rm obs}=1.5625$,
$\langle V_{x}\rangle$ drops to zero
for $30 < \theta_D < 37$ as well as
for $\theta_D > 54^\circ$
when the system enters a
jammed or clogged state for flow along
$x$, $y$, or at $45^\circ$. 
Above $r_{\rm obs} = 1.57$, flow occurs
only along the $x$ or $y$ directions,
and when $r_{\rm obs} > 3.1575$, the particle is always localized
with $\langle V_x\rangle=0$. 

\begin{figure}
\includegraphics[width=\columnwidth]{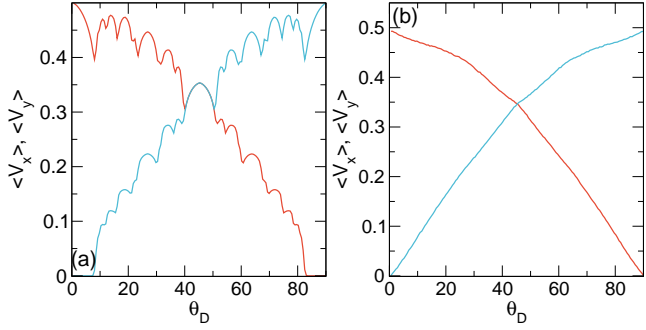}
\caption{ 
$\langle V_{x}\rangle$ (red)  and $\langle V_{y}\rangle$ (blue) vs $\theta_D$ 
for the system in Fig.~\ref{fig:11} with $F_M=0.5$, $F_D=0.5$, and
$r_{\rm obs} = 0.1$.
(a) At $l_{r} = 0.001$, there are a series of locking steps.
(b) At $l_{r} = 0.1$, the locking is lost.  
}
\label{fig:12}
\end{figure}

\begin{figure}
\includegraphics[width=\columnwidth]{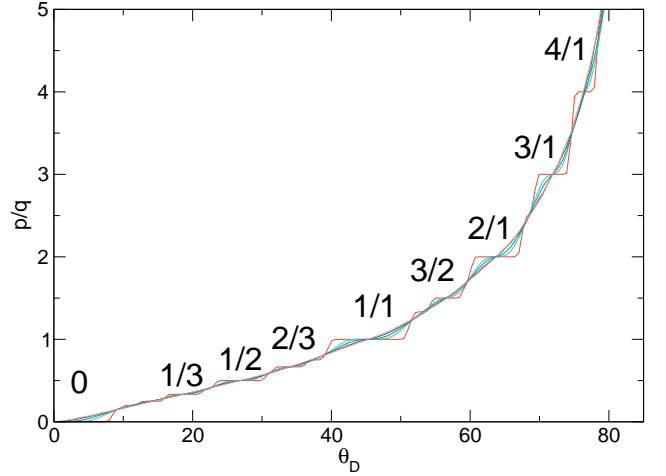}
\caption{ 
$p/q=\langle V_{y}\rangle/\langle V_{x}\rangle$ for the system in
Fig.~\ref{fig:12} with $F_M=0.5$, $F_D=0.5$, and $r_{\rm obs}=0.1$
showing steps 
at $p/q = 0$, 1/3, 1/2, 2/3, 1/1, 3/2, 2/1, 3/1, and $4/1$  
for $l_{r} = 0.001$ (dark orange), 0.01 (light blue), 0.02 (green),
0.05 (dark blue), and $0.1$ (light orange).
The locking steps gradually wash out with increasing $l_r$.   
}
\label{fig:13}
\end{figure}

In Fig.~\ref{fig:12}(a) we plot $\langle V_{x}\rangle$ and
$\langle V_{y}\rangle$ versus $\theta_D$ for the system in
Fig.~\ref{fig:11} at $l_{r} = 0.001$ and $r_{\rm obs}= 0.1$.
A series of locking steps appear, and
at the $p/q = 1/1$ step,
$\langle V_{x}\rangle=\langle V_y\rangle$.
In Fig.~\ref{fig:12}(b),
the same system at $l_{r} = 0.1$ exhibits 
smooth curves with no locking effects.
This result indicates that increasing
the activity reduces the locking
effects.
In Fig.~\ref{fig:13} we show
$p/q=\langle V_{y}\rangle/\langle V_{x}\rangle$
versus $\theta_D$ for the system in Fig.~\ref{fig:12} 
at $l_{r} = 0.001$, 0.01, 0.02, 0.05, and $0.1$,
where we highlight the steps at
$p/q = 0$, 1/3, 1/2, 2/3, 1/1, 3/2, 2/1, 3/1, and $4/1$. 
As $l_{r}$ increases, the locking steps gradually disappear.

\begin{figure}
\includegraphics[width=\columnwidth]{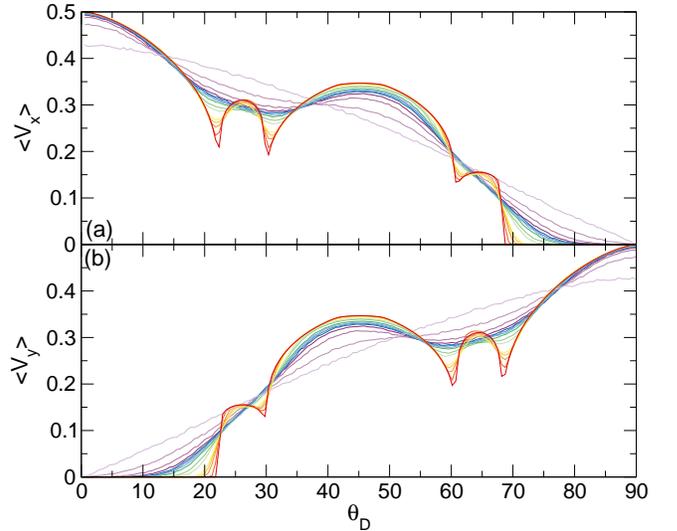}
\caption{ 
(a) $\langle V_{x}\rangle$ and (b) $\langle V_{y}\rangle$ vs
$\theta_D$ for a system with $F_{M}= 0.5$, $F_{D} = 0.5$, and
$r_{\rm obs} = 1.0$ 
at $l_{r} = 0$, 0.001, 0.003, 0.005, 0. 007, 0.02, 0.03, 0.05, 0.06,
0.07, 0.08, 0.1, 0.15, 0.3, and $1.0$, from bottom left (red) to top
left (light purple).
}
\label{fig:14}
\end{figure}

\begin{figure}
\includegraphics[width=\columnwidth]{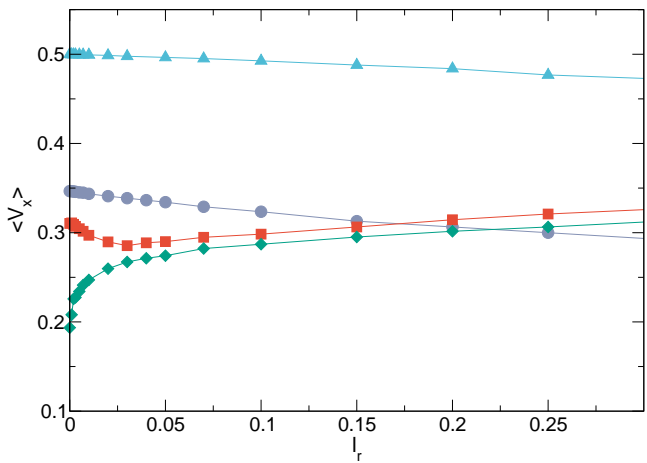}
\caption{ 
$\langle V_{x}\rangle$ measured at specific values of
$\theta_D$ vs $l_r$ for the system in
Fig.~\ref{fig:14} with $F_M=0.5$, $F_D=0.5$, and
$r_{\rm obs}=1.0$.
$\theta_D= 45^{\circ}$ (purple circles),
$\theta_D = 0^\circ$ 
(light blue triangles),
the $p/q = 1/2$ locking at $\theta_D = 26.56^\circ$ (orange squares),
and the non-locking region
at $\theta_D = 30^\circ$ (green diamonds).  
}
\label{fig:15}
\end{figure}

For obstacles of small size,
increasing the run length rapidly reduces the locking effects.
Since the locking steps become wider for larger
$r_{\rm obs}$, we next focus on samples with $r_{\rm obs} = 1.0$,
$F_{D} = 0.5$ and $F_{M} = 0.5$, which show steps for small $l_r$ at
$p/q = 0$, 1/2, 1/1, 2/1, and for $y$ direction locking.
In Fig.~\ref{fig:14}(a,b) we plot $\langle V_{x}\rangle$ and
$\langle V_{y}\rangle$ versus $\theta_D$ in this system
at $l_{r} = 0$, 0.001, 0.003, 0.005, 0.007, 0.02, 0.03, 0.05, 0.06, 0.07,
0.08, 0.1, 0.15, 0.3, and $1.0$.     
The locking steps gradually disappear as $l_r$ increases.
The $1/2$ and $2/1$ steps vanish
first, while the $1/1$ step persists up to $l_{r} = 0.3$.
At a specific value of $\theta_D$,
the values of $\langle V_{x}\rangle$ 
and $\langle V_{y}\rangle$ can decrease, increase,
or show non-monotonic behavior
as a function of $l_{r}$. 
In Fig.~\ref{fig:15} we plot
$\langle V_{x}\rangle$ versus $l_r$ at
$\theta_D= 45^\circ$ on the $p/q=1/1$ locking step,
$\theta_D= 0^\circ$ on the $p/q=0$ locking step,
$\theta_D=26.56^\circ$ on the $p/q=1/2$ locking step,
and in a nonlocking region at $\theta_D=30^\circ$.
On the $p/q = 0$ and $1/1$ locking steps,
$\langle V_x\rangle$
decreases monotonically with increasing $l_{r}$.
For $p/q = 1/2$,
$\langle V_x\rangle$ initially decreases with increasing $l_r$ until it
reaches a minimum near $l_{r} = 0.05$ and then 
increases with increasing $l_r$.
In the non-locking region at $\theta_D = 30^\circ$,
$\langle V_{x}\rangle$ monotonically increases
with increasing $l_{r}$.
Trends similar to those shown in Fig.~\ref{fig:15}
appear at higher $l_{r}$.    

\begin{figure}
\includegraphics[width=\columnwidth]{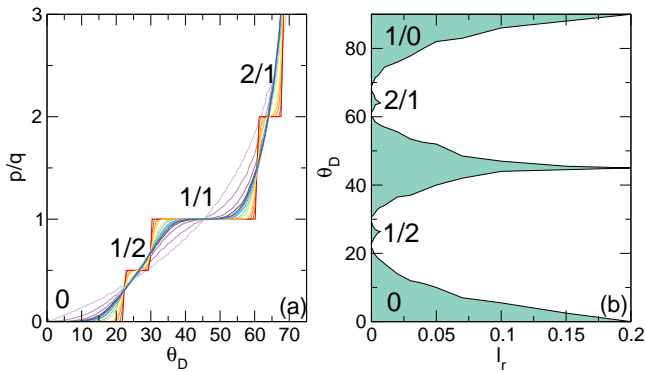}
\caption{ 
(a) The evolution of $p/q$ vs $\theta_D$ for varied $l_{r}$ for the system in
Fig.~\ref{fig:14} with $F_M=0.5$, $F_D=0.5$, and $r_{\rm obs}=1.0$
highlighting steps with $p/q = 0$, 1/2, 1/1, and $2/1$.
The value of $l_r$ ranges from $l_r=0$ (red, lower left) to $l_r=0.2$ (pale
purple, upper left).
(b) The widths of the locking steps $p/q=0$, 1/2, 1/1, 2/1, and 1/0
as a function of
$\theta_D$ vs $l_{r}$.  
}
\label{fig:16}
\end{figure}

In Fig.~\ref{fig:16}(a) we plot $p/q$
versus $\theta_D$ for the system in Fig.~\ref{fig:14}
showing that as $l_r$ increases,
the steps gradually
disappear.
This is
clearest on the $p/q=1/1$ locking step.
The fact that
the steps gradually vanish
suggests that it would be possible to 
sort particles with
varied run length $l_{r}$.
For example, if the drive angle is set to $\theta_D = 45^\circ$, particles
with 
short run lengths will lock to $45^\circ$ while particles
with longer run lengths will move at an angle
less than $45^\circ$ as
indicated in Fig.~\ref{fig:15}.
In Fig.~\ref{fig:16}(b) we illustrate the evolution
of the $p/q=0$, 1/2, 1/1, 2/1, and $y$ direction locking steps as
$l_r$ increases.
The $p/q=1/2$ and $2/1$ locking steps disappear when
$l_{r} > 0.03$, while the other locking
steps persist up to $l_{r} = 0.2$.
Partial locking (not shown) persists up to $l_r=0.3$.
For larger $r_{\rm obs}$, the
locking phases for $p/q=0$, $1/1$ and $y$ direction locking
extend out to
higher values of $l_{r}$.     

\subsection{Effect of Applied Driving Force}

\begin{figure}
\includegraphics[width=\columnwidth]{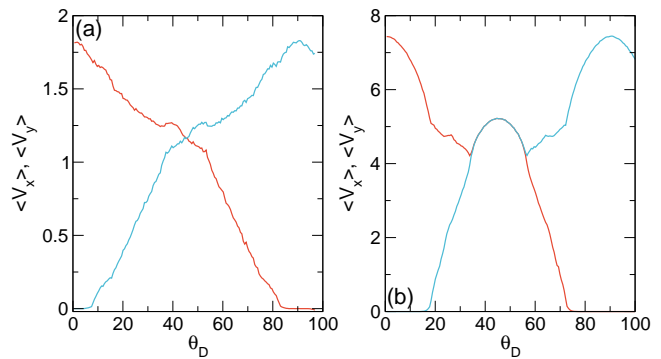}
\caption{
$\langle V_x\rangle$ (orange) and $\langle V_y\rangle$ (light blue) versus
external drive angle $\theta_D$
for a system with $F_M=0.5$, $r_{\rm obs}=1.0$, and $l_r=25$.
(a) $F_D/F_M=5.0$, where there is only a small locking window
at $p/q=1/1$.
(b) $F_D/F_M=15$, where the $p/q=1/1$ locking window is much more extended.
}
\label{fig:17}
\end{figure}

\begin{figure}
\includegraphics[width=\columnwidth]{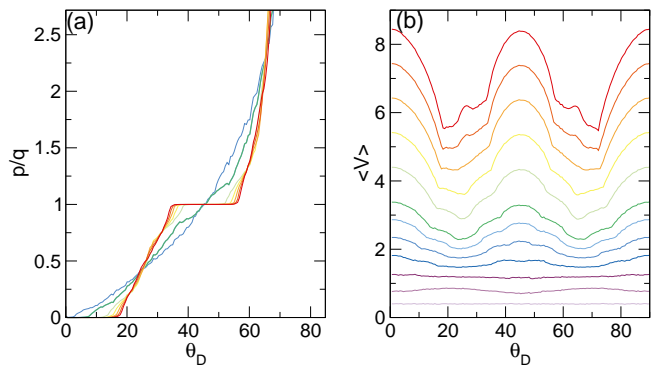}
\caption{ 
(a) $p/q$ vs $\theta_D$ for a system with $F_M=0.5$, $r_{\rm obs}=1.0$, and
$l_r=25$ at
$F_D/F_M=17$ (red), 15, 13, 11, 9, 7, 6, 5, 4, and 3 (blue), from
lower left to upper left, showing that the $p/q=1/1$ step disappears
as $F_M/F_D$ decreases.
(b) $\langle V\rangle$ vs $\theta_D$
in the same system at
$F_D/F_M=17$ (red), 15, 13, 11, 9, 7, 6, 5, 4, 3, 2, 1.5, and 1 (light purple),
from top to bottom.
}
\label{fig:18}
\end{figure}
      
\begin{figure}
\includegraphics[width=\columnwidth]{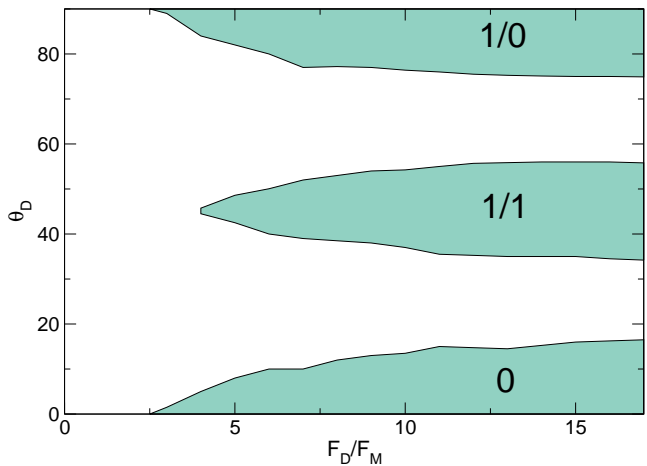}
\caption{
Locations of the $p/q=0$, 1/1, and 1/0 steps as a function of
$\theta_D$ versus $F_D/F_M$ for a system with
$F_M=0.5$, $r_{\rm obs}=1.0$, and $l_r=25$.
}
\label{fig:19}
\end{figure}

We next consider a system in which both the direction and the magnitude
of the external biasing force are varied while the motor force
magnitude $F_M$ is held fixed.
We consider a range of values from the motor force
dominated regime with $F_{D}/F_{M} \leq 1.0$
to the external biasing force dominated regime with
$F_{D}/F_{M} > 1.0$.
As shown in the previous section,
when $F_{D}/F_{M} \approx 1.0$ the 
directional locking effects disappear
for $l_{r} \geq 1.0$.
We
consider samples with $F_{M} = 0.5$ and 
$l_{r} = 25$, a combination which produces no locking in any direction
when $F_{D}/F_{M} < 2.0$.
In Fig.~\ref{fig:17}(a) we
plot
$\langle V_{x}\rangle$
and $\langle V_{y}\rangle$
at $F_{D}/F_{M} = 5.0$,
where locking in $x$ and $y$ appears in only a small region
with $p/q = 1/1$. 
At $F_{D}/F_{M} = 15$ in Fig.~\ref{fig:17}(b),
an extended $p/q=1/1$ locking region appears over which
$\langle V_x\rangle=\langle V_y\rangle$.
In Fig.~\ref{fig:18}(a) we plot $p/q$ versus $\theta_D$
for the system in Fig.~\ref{fig:17} 
at $F_{D}/F_{M} = 17$, 15, 13, 11, 9, 7, 6, 5, 4, and $3.0$,
highlighting the $p/q=0/1$ and $1/1$ locking steps.
For $F_{D}/F_{M} = 3.0$, there is almost no locking at $p/q=0/1$ and no
locking at $p/q=1/1$, while
for $F_{D}/F_{M} = 4.0$, weak locking appears at $p/q=0/1$ but
there is still no locking at $p/q=1/1$.
As $F_{D}/F_M$ increases, the $p/q=1/1$ locking step emerges and becomes
wider, and indications of partial locking appear
near $p/q=1/4$ and $p/q=1/2$.
In Fig.~\ref{fig:18}(b) we plot
the total velocity $\langle V\rangle$ versus $\theta_D$
for the same system at
$F_{D}/F_{M} = 17$, 15, 13, 11, 9, 7, 6, 5, 4, 3, 2, 1.5, and $1$.
As $F_{D}/F_M$ increases, the net velocity increases,
and strong oscillations emerge in $\langle V\rangle$ so that
the locking steps at $p/q=0/1$, 1/1, and
$y$ direction locking appear as clear bumps.
For $F_{D}/F_{M} = 17$ and $15$, there are
smaller bumps near $\theta = 26^\circ$ and $65^\circ$,
corresponding to $p/q=1/2$ and $p/q=3/2$; however,
this locking is only partial.
For $F_{D}/F_{M} = 1.0$, $\langle V\rangle$ does not depend on $\theta_D$,
indicating the absence of full or 
partial locking.
In Fig.~\ref{fig:19} we plot the locations of the
$p/q=0$, 1/1, and 1/0 steps as a function of
$\theta_D$ versus $F_{D}/F_{M}$,
showing that locking effects occur only when $F_{D}/F_{M} > 2.0$. 
These results indicate that even for very long run lengths,
directional locking can occur
as long as the external biasing force
is at least twice as large as the motor force.
This suggests that particles that couple differently to the external
drive could be separated,
with slow moving active particles following the direction of the
external drive and
fast moving active particles locking to one of the
different symmetry directions. 

\subsection{Clogging Effects}

\begin{figure}
\includegraphics[width=\columnwidth]{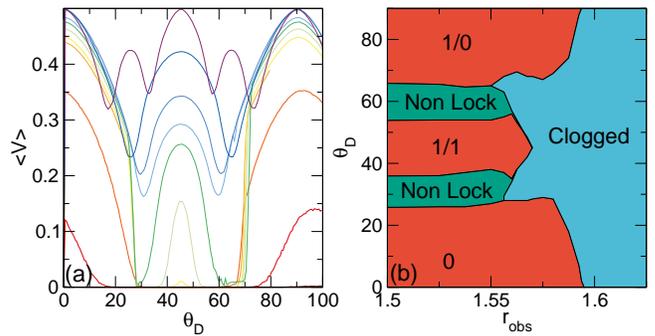}
\caption{ 
(a) The total velocity $\langle V\rangle$ vs $\theta_D$
for a system with $F_{D} = 0.5$, $F_{M} = 0.5$, and $l_{r} = 0.01$ for 
$r_{\rm obs} = 0.75$, 1.25, 1.5, 1.55, 1.556, 1.562, 1.5675, 1.57, 1.58,
1.5875, and $1.5925$, from center top (purple) to center bottom (red).   
(b) Dynamic phase diagram as a function of $\theta_D$ vs $r_{\rm obs}$
for the system in (a) with $r_{\rm obs} > 1.5$, showing the locations
of the $p/q=0$, 1/1, and 1/0 locking steps, non-locking regimes,
and the clogged state.
}
\label{fig:20}
\end{figure}

We next consider large obstacles which can induce
clogging effects depending on the
level of activity in the systems.
In Fig.~\ref{fig:20}(a) we plot the total velocity $\langle V\rangle$
versus $\theta_D$ 
for a system with $F_{D} = 0.5$, $F_{M} = 0.5$,
and $l_{r} = 0.01$ at $r_{\rm obs} = 0.75$, 1.25, 1.5, 1.55, 1.556, 1.562,
1.5675, 1.57, 1.58, 1.5875, and $1.5925$.
For $r_{\rm obs} = 0.75$, locking steps appear at $p/q=0$, $1/2$, $1/1$,
$3/2$, and $y$ direction locking.
As $r_{\rm obs}$ increases, the
net velocity decreases.
When $r_{\rm obs} < 1.562$,
$\langle V\rangle$ drops to zero for driving
in certain directions but locking
steps are still present for $\theta_D=0^\circ$, $45^\circ$, and $90^\circ$,
while for $r_{\rm obs} > 1.57$ locking
occurs only along the $x$ and $y$ directions.
At large enough $r_{\rm obs}$, the system becomes completely clogged.
In Fig.~\ref{fig:20}(b) we show a dynamic phase diagram
for the system in Fig.~\ref{fig:20}(a) in the regime $r_{\rm obs}>1.5$, where
we highlight
the $p/q=1/1$ and 1/0 locked phases, regions where unlocked flow occurs,
and the clogged state.
The clogging regions interdigitate with
the non-locking regions, indicating
that there is a higher susceptibility to clogging
for driving at certain non-locking angles.
For the larger obstacle sizes, the system is always 
clogged.

The tendency for a system with a periodic array of obstacles
to clog
for driving in different
directions was previously studied
for passive bidisperse disks in Ref.~\cite{Nguyen17}.
Even when free flow of the disks is possible along the
$x$ or $y$ directions, the system can clog
for driving at incommensurate angles due to collisions between the disks
and the obstacles.
For a square obstacle array, the disks can move without collisions
for driving along $45^\circ$;
however, the effective distance between the obstacles is smaller compared to
driving in the $x$ or $y$ directions, so the system reaches a clogged
state at lower $r_{\rm obs}$ for the $45^\circ$ driving.
In the active particle system we consider here,
the clogging is a single
particle effect that is produced by the interplay between
the drive force, the motor force, and the disk-obstacle interactions.
For a disk-obstacle interaction with a softer form,
the onset of clogging shifts to
larger values of $r_{\rm obs}$,
whereas if $F_{D}$ or $F_{M}$ are reduced, the clogging onset
shifts to smaller values of $r_{\rm obs}$.
The amount of clogging
that occurs for a fixed drive and motor force also depends on
the size of $l_{r}$. If collective
particle-particle interactions become important,
distinct types of jamming or clogging effects
\cite{Liu98,Zuriguel15,Barre15,Stoop18} could arise
that differ from what we observe.
The activity could reduce clogging effects
at small run lengths, but could increase the clogging or
induce
partial clogging for large run lengths,
as shown in studies of active particles driven through
randomly placed obstacles \cite{Sandor17a,Reichhardt18c}.       

\begin{figure}
\includegraphics[width=\columnwidth]{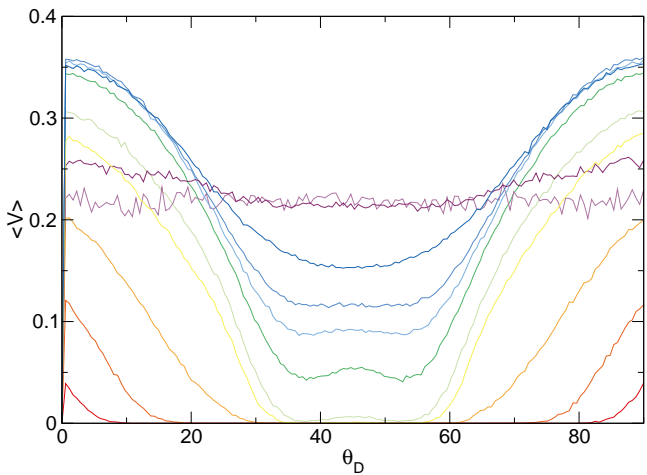}
\caption{ 
$\langle V\rangle$ vs $\theta_D$ for the system in
Fig.~\ref{fig:20} with $F_D=0.5$, $F_M=0.5$, and $r_{\rm obs} = 1.5875$ at
$l_{r} = 0.005$, 0.01, 0.03, 0.07, 0.1, 0.2, 0.3, 0.4, 0.7, 5, and $25$,
from center bottom (red) to center top (light purple). 
}
\label{fig:21}
\end{figure}

In Fig.~\ref{fig:21} we plot
$\langle V\rangle$ versus $\theta_D$ for the system in
Fig.~\ref{fig:20} at $r_{\rm obs} = 1.5875$ for
$l_{r} = 0.005$, 0.01, 0.03, 0.07, 0.1, 0.2, 0.3, 0.4, 0.7, 5, and $25$.
Increasing the run length produces a variety of effects.
At $\theta_D = 0^\circ$,
$\langle V\rangle$ gradually increases with increasing $l_r$
until reaching the value $\langle V\rangle = 0.355$
near $l_{r} = 0.7$. As $l_r$ is increased further,
$\langle V\rangle$ decreases
to $\langle V\rangle = 0.21$ due to a self-trapping effect.
At $\theta_D = 45^{\circ}$,
$\langle V\rangle$ monotonically increases with increasing $r_{l}$ and the
clogging effect disappears entirely
for $l_{r} > 0.1$.
If $r_{\rm obs}$ is increased,
the value of $l_r$ at which an
unclogged state appears increases.
A local maximum in $\langle V\rangle$ appears
at $\theta_D = 45^\circ$ when $l_{r}= 0.1$, 0.2, and $0.3$
due to a partial locking to
the $p/q = 1/1$ step.
These results suggest that active particle sorting could be achieved
using clogging effects in which
the less active particles with shorter run lengths
would be trapped but
the more active particles with longer run lengths would be mobile.    

\begin{figure}
\includegraphics[width=\columnwidth]{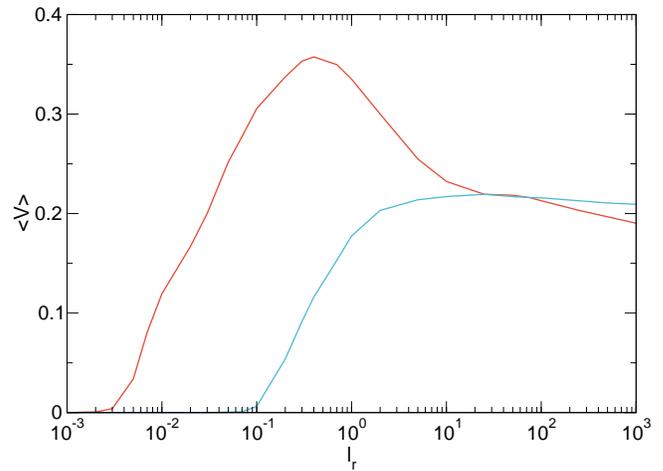}
\caption{  
$\langle V\rangle$ vs $l_{r}$
at $\theta_D=0$ (orange) and $\theta_D=45^\circ$ (blue)
for the system in
Fig.~\ref{fig:21}
with $F_D=0.5$, $F_M=0.5$, and $r_{\rm obs}=1.5875$.
}
\label{fig:22}
\end{figure}

In Fig.~\ref{fig:22}
we plot $\langle V\rangle$ versus $l_r$ for the system in
Fig.~\ref{fig:21} at $\theta_D = 0^\circ$ and $\theta = 45^\circ$.
For $l_r < 0.01$, the system is completely clogged.
When $l_{r} \geq 0.01$, $\langle V\rangle$
for $\theta_D=0^\circ$ increases with increasing
$l_r$ to a maximum value
near $l_{r} = 1.0$, after which $\langle V\rangle$ decreases with
increasing $l_r$.
At $\theta_D = 45^\circ$, the sample
is clogged below $l_{r} = 0.5$.
There is then an increase in $\langle V\rangle$ with increasing
$l_r$ up to 
a maximum value near $l_{r} = 100$, followed by a slight
decrease in $\langle V\rangle$.

\section{Summary}
We have examined active run and tumble particles interacting with a square
array of obstacles. For a system without
any external biasing,
we find that for short run times the particles
act close to the Brownian limit and explore 
space randomly.
For long run times, the particles become directionally locked and
move only along certain symmetry directions of the substrate.
These directions correspond to angles
$\theta = \arctan(p/q)$ where $p$ and $q$ are integers.
As the radius of the obstacles increases,
the number of locking angles decreases until
only the steps at $p/q = 0$, $p/q=1/1$, and $y$ direction locking remain.
The locking can be measured by
examining the ratio of the $x$ and $y$ direction velocities
as well as the instantaneous velocity distribution functions. 
When an additional biasing drive is applied in the $x$-direction,
the average drift velocity decreases if the run time is increased,
while an increase in the magnitude of the external biasing
force relative to the motor
force can produce peaks in the differential
velocity-force curves.
As the direction of the external drive is changed,
we observe a directional
locking effect similar to that found in non-active systems.
Increasing the run time destroys the directional locking, but
if the ratio of the biasing driving force to the
motor force is made large enough, the directional locking reappears.
For large obstacles, the
system can exhibit a directional depinning
and clogging effect in which
increasing the run time can induce the onset of clogging.

\begin{acknowledgments}
We gratefully acknowledge the support of the U.S. Department of
Energy through the LANL/LDRD program for this work.
This work was supported by the US Department of Energy through
the Los Alamos National Laboratory.  Los Alamos National Laboratory is
operated by Triad National Security, LLC, for the National Nuclear Security
Administration of the U. S. Department of Energy (Contract No. 892333218NCA000001).

\end{acknowledgments}

\bibliography{mybib}
\end{document}